\documentstyle[aps,multicol,prl,epsfig,floats]{revtex}
\begin{document}
\draft
\twocolumn[\hsize\textwidth\columnwidth\hsize\csname @twocolumnfalse\endcsname
\author{Yu. B. Khavin and M. E. Gershenson}
\address{Rutgers University, Department of Physics and Astronomy, \\Serin Physics
Laboratories, Piscataway, NJ 08854-8019}
\author{A. L. Bogdanov}
\address{Lund University, MAX-lab, National Laboratory, \\
S-221 00 Lund, Sweden}
\date{\today }
\title{Strong localization of electrons in quasi-one-dimensional conductors}
\maketitle

\begin{abstract}
We report on the experimental study of electron transport in sub-micron-wide
''wires'' fabricated from Si $\delta $-doped GaAs. These quasi-one-dimensional
(Q1D) conductors demonstrate the crossover from weak to strong
localization with decreasing the temperature. On the insulating side of the
crossover, the resistance has been measured as a function of temperature,
magnetic field, and applied voltage for different values of the electron
concentration, which was varied by applying the gate voltage. The activation
temperature dependence of the resistance has been observed with the
activation energy close to the mean energy spacing of electron states within
the localization domain. The study of non-linearity of the current-voltage
characteristics provides information on the distance between the critical
hops which govern the resistance of Q1D conductors in the strong
localization (SL) regime. We observe the exponentially strong negative
magnetoresistance; this orbital magnetoresistance is due to the universal
magnetic-field dependence of the localization length in Q1D conductors. The
method of measuring of the single-particle density of states (DoS) in the
 SL regime has been
suggested. Our data indicate that there is a minimum of DoS at the Fermi
level due to the long-range Coulomb interaction.
\end{abstract}

]

\section{INTRODUCTION}

Recent progress in technology enables the realization of a wide variety of
materials with one-dimensional (1D) structural and electronic properties:
high-mobility heterojunction microstructures \cite{fow1}, heavy-doped
conjugated polymers \cite{ish}, carbon nanotubes \cite{ebb}, or organic
conductors \cite{voit}, to mention a few. Because of a very broad current
usage of the term ''1D systems'', the physical properties of these
conductors are diverse. In the limit of one conducting channel (conductors
with cross-sectional dimensions smaller than the Fermi wavelength 
of the conduction electrons), there is strong unscreened interaction
between electrons. The electron states are correlated along the channel, and
the poorly-defined single-electron excitations cannot be treated as Landau
quasiparticles. The behavior of these so-called quantum ''wires'' is
described by the Tomonaga-Luttinger model, and the ideas span from the
Wigner crystal in the case of the long-range interactions to the
charge-density waves for the short-range interactions (for recent reviews,
see \cite{voit,mau,sch,gog}).

In this paper, we are concerned with another class of 1D conductors, usually
referred to as quasi-one-dimensional (Q1D) conductors. In these disordered
conductors, there are many channels with strong scattering between them,
and the quasiparticle excitations are well described by the Fermi-liquid
theory. The physics of these systems is essentially different from the
physics of quantum wires. The electron mean free path $l$ is much smaller
than the length of these conductors, and the coherent scattering from many
impurities gives rise to Anderson localization\cite{thoul}. To be
one-dimensional with respect to the quantum interference effects, a
conductor should have cross-sectional dimensions smaller than the length of
localization of the electron wavefunction, $\xi $, and the phase-breaking
length $L_\varphi $ (for a review, see\cite{aarev}).

It is widely believed that all electron states in low-dimensional conductors
are localized when both spin-orbit (SO) and electron-electron interactions
are weak\cite{thoul,abrahams}. The localization length for a Q1D conductor
with a large number of transverse channels $N_{1D}$ and weak SO scattering
can be expressed at $H=0$ as \cite{larkin,imry} 
\begin{equation}
\xi =N_{1D}l=\frac{\pi \hbar }{e^2}\sigma _1\text{ ,}  \label{ksi}
\end{equation}
where $\sigma _1$ is the conductance of a wire
per unit length in the ''metallic'' regime. Despite of electron
localization, this ''metallic'' conductance can be rather large at room
temperature. This is due to strong inelastic scattering: the electron
scatters to another state, localized around a different site, before it
diffuses over the localization length [the weak localization (WL) regime].
However, with decreasing the temperature, a low-dimensional conductor should
eventually become an insulator. Electron transport can proceed only by
hopping in this strong localization (SL) regime.

Until recently, the temperature-driven crossover has been observed only in
two-dimensional conductors\cite{IO,white,jiang,hsu}. In one-dimensional
MOSFET-type structures, the transition to the insulating regime has been
observed with decreasing the carrier concentration\cite{fow1}. In this case,
however, all electron parameters {\it and} disorder have been changed
simultaneously with variation of the gate voltage, and the electron states
were {\it quite} {\it different} in the ''metallic'' and insulating regimes.

The study of the crossover is more informative if the crossover is observed
as a function of temperature: in this case the data obtained in the WL and
SL regimes are pertinent to the {\it same} electron states. Though the
theoretical prediction of this remarkable crossover in Q1D conductors was
made by Thouless in 1977\cite{thoul}, the experimental study of this
fundamental problem was delayed for 20 years. The ''gap'' between the
prediction and observation indicates that this is a very demanding
experiment; in particular, the choice of adequate samples is crucial. 
Two objects which have been used extensively for the study of the
WL regime, thin metal films and high-mobility heterostructures, do not suit
well this purpose: the localization length in these conductors is too large,
and, hence, the crossover temperature is too low for any conceivable
cross-section\cite{thorn,mani}.

Experiments \cite{thorn,prob,gersh0} have demonstrated
that the dominant decoherence mechanism in Q1D conductors at low
temperatures is the quasi-elastic electron-electron scattering\cite{aak}.
Analyzing these data, we came to a conclusion that the crossover temperature
can be substantially increased by using low-mobility and heavy-doped
semiconductor structures. Recently we observed for the first time the {\it %
temperature-driven} crossover in sub-micron-wide ''wires'' fabricated from
the Si $\delta $-doped GaAs\cite{gersh1,gersh2,gersh3}. On the insulating
side of the crossover, there remain unanswered questions that are crucial
for understanding of the transport mechanisms in Q1D conductors. In this
paper, we focus on the study of the conductivity of Q1D conductors in the
strongly localized regime.

The paper is organized as follows. In Section II, we briefly describe the
experimental technique and observation of the WL-SL crossover in our
samples. The data obtained in the SL regime are discussed in Section III.
Our experimental findings and conclusions are summarized in Section IV.

\section{OBSERVATION OF THE WL-SL CROSSOVER}

\subsection{Samples}

One can estimate the crossover temperature $T_\xi $ using the Thouless'
idea that the localization length $\xi $ and
the phase-breaking length $L_\varphi (T_\xi )$ should be of the same order
of magnitude at the crossover\cite{thoul}. At low temperatures, the decoherence in Q1D
conductors is due to the quasi-elastic electron-electron scattering (the
so-called Nyquist decoherence mechanism)\cite{aak}. The phase-breaking
length in this case can be expressed as follows\cite{aak}: 
\begin{equation}
L_\varphi =\left( \frac{\hbar ^2D\sigma _1}{\sqrt{2}e^2k_BT}\right) ^{1/3}.
\label{Lfai}
\end{equation}
From comparison Eqs.\ref{ksi} and \ref{Lfai}, we obtain the following
estimate 

\begin{equation}
T_\xi =\frac{e^2D}{\sqrt{2}\pi ^2k_B\xi \sigma _1}\sim \left[ DW^2\left(
m^{*}\right) ^2\right] ^{-1}\text{ ,}  \label{Tksi}
\end{equation}
where $D$ is the electron diffusion constant in the metallic regime,
$W$ is the width of a "wire" fabricated from two-dimensional electron
gas, and $m^{*}$ is the effective electron mass.
Thus, one could expect larger $T_\xi $ for {\it narrow disordered}
conductors with a {\it small} effective mass of the current carriers.

We have used wires fabricated from $\delta $-doped GaAs. The effective mass
of electrons in these structures coincides with $m^{*}\simeq 0.067{\it m}_{%
\text{e}}$ in GaAs. A single $\delta $-doped layer 
with concentration of Si donors $N_D=5\cdot 10^{12}cm^{-2}$ is $0.1$ $%
\mu m $ beneath the surface of an undoped GaAs. Using the electron beam
lithography and deep ion etching, we were able to prepare uniform conducting
wires with the effective width $W$ as small as $0.05$ $\mu m$. Because of
the side-wall depletion, the effective width is smaller than the
''geometrical'' one by $0.15\div 0.2$ $\mu m$, depending on the carrier
concentration. The values of $W$, obtained from the sample resistance, were
in accord with the estimate of $W$ from the analysis of the WL
magnetoresistance. Parameters of the samples are listed in Table 1.

For several samples, we repeated the whole set of measurements after
deposition of a thin ($\sim 50nm$) silver film on top of the structure. This
metal film was used as a gate electrode: by varying the gate voltage $V_g$
, we could ''tune'' the carrier concentration and
mobility, and, hence, the localization length. The metal film deposition
also serves a different purpose: it modifies the Coulomb interaction at
distances greater than the distance between the electron gas and the
metal film $t=0.1\mu m$, and allows to test the effect of the
electron-electron interactions on the conductivity in the SL regime.

The carrier concentration $n$ in the wires could differ
substantially from that in the 2D $\delta $-doped layer because of the
side-wall depletion. The direct measurement of $n$ (e.g., from the
Shubnikov-de Haas oscillations) cannot be performed in these very disordered
and narrow wires (see, e.g. \cite{berg}). An indirect estimate of $n$ can
be done as follows. It is well established that the electron mobility $\mu $
in the $\delta $-doped layers with $N_D=5\cdot 10^{12}$cm$^{\text{-2}}$ is $%
\sim (1\pm 0.2)\cdot 10^3$ $cm^2/V\cdot s$ \cite{mobility}. Assuming that $%
\mu =10^3cm^2/V\cdot s$ in our samples, we obtain a reasonable estimate $%
n\approx 3\cdot 10^{12}$cm$^{-2}$ for $V_g$ $=0$; this corresponds to $\sim
40\%$ compensation of Si donors, which is typical for the $\delta $-doped
layers with $N_D=5\cdot 10^{12}$cm$^{\text{-2}}$ (see, e.g. \cite{compens}).
For this carrier concentration, only the lowest 2D subband is occupied \cite
{2Dsub}. It is worth noting that the knowledge of the exact value of $n$ 
{\it is not crucial} for most of the effects discussed below.

A relatively high concentration of carriers ensures that the number of
transverse channels in our samples is large ($N_{1D}$ $\approx 7-30$). The
localization length is always much greater than the mean free path $l$, 
and electron motion is 
{\it diffusive} within the localization domain. At high temperatures, these
samples can be considered as a disordered two-dimensional metal with the
Fermi energy $\varepsilon _F$ of the order of $\sim10^3K$, and the parameter $k_Fl$
ranging from $6$ to $40$ ($k_F$ is the Fermi wave number). In particular,
the density of states for non-interacting electrons should be energy-independent,
as in two dimensions, because of the strong inter-channel scattering ($\hbar
/\tau >>\varepsilon _F/N_{1D}$, where $\tau $ is the momentum relaxation time).
However, at low temperatures the samples become {\it one-dimensional} with
respect to the quantum interference effects : $W<L_\varphi (T)\leq \xi $.

\bigskip

Table 1. Parameters of the samples. \bigskip

\begin{tabular}{|c|c|c|c|c|c|c|}
\hline
Sample, \# & 1 & 2 & 3 & 4 & 5 & 6 \\ \hline
$W$, $\mu m$ & 0.05 & 0.06 & 0.1 & 0.12 & 0.2 & 0.18 \\ \hline
$L,\mu m$ & 500 & 500 & 40 & 500 & 40 & 500 \\ \hline
\# of parallel ''wires'' & 470* & 470 & 5 & 470 & 5 & 470 \\ \hline
$R_{\Box }(T=20K),k\Omega $ & 1.6 & 1.7 & 3.5 & 1.6 & 4.2 & 1.7 \\ \hline
$\xi ,\mu m$ & 0.40 & 0.46 & 0.37 & 1.0 & 0.61 & 1.4 \\ \hline
$\Delta _\xi ,K$ & 2.1 & 1.5 & 1.1 & 0.35 & 0.34 & 0.17 \\ \hline
$T_0(H=0),K$ & 2.6 & 1.87 & 1.47 & 0.42 & 0.39 & 0.2 \\ \hline
$R_\xi (T=T_0),k\Omega $ & 20.4 & 21.3 & 28 & 23 & 24.4 & 24.3 \\ \hline
$H_\xi ,kOe$ & 1.0 & 0.74 & 0.56 & 0.17 & 0.17 & 0.083 \\ \hline
$H_\xi ^{exp},kOe$ & 1.0 & 0.80 & 0.51 & 0.21 & 0.17 & 0.12 \\ \hline
$H_\xi ^{exp}/T_0,kOe/K$ & 0.37 & 0.43 & 0.35 & 0.50 & 0.44 & 0.59 \\ \hline
\end{tabular}

\bigskip

*) sample 1 has been scratched during the gate deposition, and after this
it contained 360 wires.

\bigskip

In these samples, the electrons are localized over a large area $\xi \cdot W$,
which is shared by several thousand of the other localized electrons. Though
the electron states strongly overlap in space, they are separated by the
mean energy spacing within the localization domain

\begin{equation}
\Delta _\xi (1D)=\left( \nu _{1D}\xi \right) ^{-1}\text{ ,}
\end{equation}
where $\nu _{1D}$ is the single-particle density of states (due to the strong
inter-channel scattering, $\nu _{1D}=W\nu _{2D}$).
For the samples discussed in this paper, this energy spacing varies from $0.1K$ to $%
5K$ depending on the wire width and the carrier concentration.

We studied the resistance of many wires connected in parallel. This has been
done for two reasons. First, the parallel connection of wires enlarges the
temperature interval where the sample resistance is within the range of our
measuring equipment ($\leq 1G\Omega $). Second, increasing the number of
wires and their length reduces mesoscopic fluctuations by improving the
ensemble averaging. Initially, we studied five $40\mu m$-long wires
connected in parallel \cite{gersh1}, later the number of wires was increased
up to 470, and the length $L$ of each wire - up to $500\mu m$\cite
{gersh2,gersh3}. The distance between the wires is $1\mu m$ in all the
samples. The longer wires have an additional important advantage: the
voltage interval that corresponds to the {\it linear} current-voltage ($I$-$V
$) characteristics is narrow for our samples (see Sec. III), and the use of
the longer wires facilitates measurements in the linear regime.

\subsection{Measuring technique}

For the measurements, we use a dilution refrigerator with a base temperature
of $30mK$. All the wiring in the refrigerator is done by shielded twisted
pairs of wires; the resistance between all the wires is much greater than $%
1G\Omega $. An external noise pick-up has been minimized by filtering of all
the wires going into the cryostat.

We exploit two techniques for measuring the resistance. For $R\leq 5M\Omega $%
, we have used a lock-in amplifier with an input resistance of $20M\Omega $;
a low measuring frequency ($f=0.5Hz$) has been chosen because of large
values of the sample resistance and capacitance of the filters. Due to a
very high sensitivity of the lock-in technique, high-resolution resistance
measurements can be done at low voltage levels ($V\sim 10^{-6}V$). This is
an important advantage of the ac measurements, because the region of
linearity of the $I-V$ characteristics becomes narrower with increasing the
localization length: e.g., for a sample with $\xi \sim 1\mu m$ and $L=500$ $%
\mu m$ the non-linearity is observed at $V$ as small as 10$^{-4}$ $V$ (see
Sec. III). For the resistance measurements in the range $10M\Omega \leq
R\leq 1G\Omega $, we have used a dc current source and electrometer with the
input resistance greater than $1\cdot 10^{14}$ $\Omega $. The electrometer
resolution, which was mostly limited by slow drifts, has been increased up
to $\Delta V$ $\sim 1\cdot 10^{-5}$ $V$ by alternating of the measuring
current direction.

\subsection{Crossover from weak to strong localization}

The resistance of all the samples increases with decreasing the temperature
(Fig. 1). At high temperatures, a slow growth of $R$ is consistent with the
theory of quantum corrections to the resistance in the WL regime \cite
{gersh3}. However, a dramatic change in the temperature dependence of the
resistance has been observed for sufficiently narrow ($W$ $<$ 0.3 $\mu $m)
wires: it becomes activation-type at low temperatures.

We have shown that the Q1D conductors are driven into the insulating state
by {\it both} single-particle localization and electron-electron interaction%
\cite{gersh3}. This evidence stems from the study of precursors of the
crossover, namely, from the quantitative analysis of the temperature and
magnetic field dependences of the resistance on the ''metallic'' side of the
crossover. The contributions to the resistance due to localization and
interaction effects are of the same order of magnitude at the crossover. The
temperature dependence of the resistance is well described by the sum of the
first-order quantum corrections down to $\sim3T_\xi $, where $T_\xi $
is the crossover temperature; at lower temperatures, the higher-order
corrections become important (see Section IIIA for our method of finding $%
T_\xi $).

Since our samples are driven into the insulating state by both localization
and interaction effects, it is not obvious that the Thouless' scenario,
which has been suggested for a system of non-interacting electrons \cite
{thoul}, applies in this case. However, the experiment demonstrates
that the features of the observed crossover are consistent with the Thouless
predictions. 

First, the resistance $R_\xi $, calculated for a wire segment
of a length of $\xi $, is 24$\pm $4 k$\Omega $ at the crossover temperature
(see Table 1 and Fig.1), which is close to the quantum resistance $h/e^2$
expected for such a wire in the vicinity of the Thouless crossover.

Secondly, the theory \cite{thoul} predicts that the crossover occurs when
the temperature-dependent phase-breaking length $L_\varphi (T)$ becomes of
the order of the temperature-independent localization length. We estimated $%
L_\varphi (T)$ by fitting the high-temperature $(T>T_\xi )$
magnetoresistance with the WL theory \cite{gersh3} (for a detailed procedure
of fitting the experimental magnetoresistance of Q1D conductors with the
theory\cite{aak}, see Ref.\cite{gersh0,gersh3}). It has been shown that the
dominant phase-breaking mechanism in our samples on the metallic side of the
crossover is the quasi-elastic electron-electron scattering 
\cite{aak}; both the temperature dependence of the
phase-breaking length ($L_\varphi \sim T^{-1/3}$) and its magnitude are in a
good agreement with the theoretical result (2) \cite{gersh3}. For all the
samples studied, $L_\varphi $ at the crossover temperature is approximately
2-3 times smaller than $\xi $ calculated from Eq. 1. However, it would be
naive to expect the exact equality $L_\varphi (T_\xi )=\xi $ at $T=T_\xi $,
since the prediction has an ''order-of-magnitude'' character.

The good agreement with Thouless' prediction indicates that in our samples the 
localization effects prevail in the WL-SL crossover. An additional evidence 
for that is provided by observation of the exponentially strong orbital
magnetoresistance in the SL regime and the decrease of the crossover
temperature in classically weak magnetic fields (see Sec. IIIC).

Thus, we observe for the first time the crossover from weak to strong
localization in Q1D conductors as a function of temperature, when the
electron states and disorder are {\it identical} on either side of the
crossover. This important aspect of the experiment allows us to compare
information on the electron states which can be obtained independently from the
study of conductivity on {\it both sides} of the crossover.

\section{THE STRONG LOCALIZATION REGIME}

\subsection{The temperature dependence of the resistance}

On the insulating side of the crossover, an activated behavior of the
resistance is observed (see Fig. 1): the experimental dependences $R(T)$ at $%
T\leq 0.3$ $T_0$ can be fitted with the Arrhenius-type dependence 
\begin{equation}
R(T)=R_0\exp \left( \frac{T_0}T\right) \text{ .}  \label{Arr}
\end{equation}
Similar $R(T)$ dependences were reported earlier for the mesoscopic
MOSFET-type structures in the SL regime \cite{fow1}. 

\begin{figure}[th]
\epsfig{file=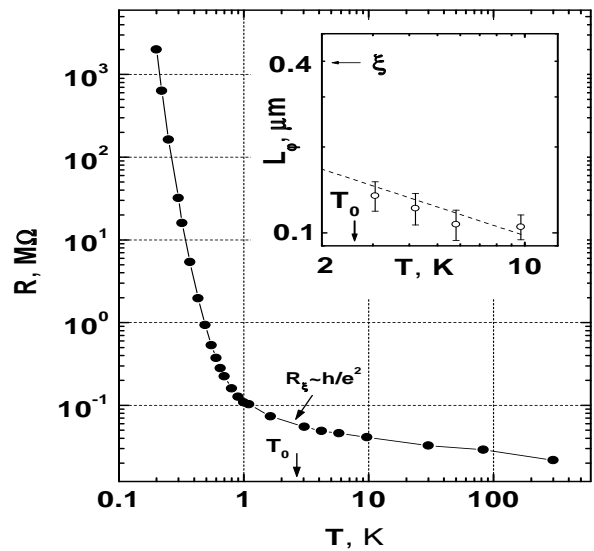, height=0.5\textwidth, width=0.5\textwidth}
\caption{The temperature dependence of the resistance of the 0.05-$\mu$%
m-wide wires (sample 1) in zero magnetic field without the gate, the
solid curve is a guide to the eye. The arrow indicates the temperature $T_0$
that corresponds to the activation energy of hopping transport on the insulating
side of the crossover. The insert: the temperature dependence of the
phase-breaking length $L_\varphi$. The dashed line is the Nyquist phase-breaking 
length (Eq. 2). }
\label{Fig.1}
\end{figure}

The experimental values of 
the activation energy $k_BT_0$ are very close to the spacing of the
electron states on the scale of the localization domain $\Delta _\xi $. We
have verified this a) by varying the width of samples, and b) by varying the
localization length with the gate voltage. In particular, Table 1
demonstrates that $T_0$ for the samples with the same diffusion constant
varies proportional to $W^{-2}$, as one could expect from Eq. \ref{Tksi}.

At the crossover temperature, {\it all relevant energy scales} become of the
same order of magnitude: 
\begin{equation}
\frac \hbar {\tau _\varphi (T_\xi )}\sim \Delta _\xi \sim \frac{\hbar D}{\xi
^2}\sim k_BT_0\sim k_BT_\xi \text{ .}  \label{allenergies}
\end{equation}
Indeed, the scaling theory of localization\cite{thoul,abrahams} predicts
that the crossover occurs when the smearing of the energy levels  $\hbar
/\tau _\varphi $ becomes comparable with the level spacing within the
localization domain $\Delta _\xi $. For a Q1D conductor, $\Delta _\xi $ is
of the order of the Thouless energy, $\hbar D/\xi ^2$ (Eq. \ref{ksi}).
Our experimental data indicate that the activation energy $k_BT_0$ is also
very close to the level spacing $\Delta _\xi $ (see below). Finally, if the Nyquist
phase-breaking is the dominant decoherence mechanism (which is always the
case in low-dimensional conductors at sufficiently low temperatures), the
phase-breaking rate $\hbar /\tau _\varphi $ becomes of the order of the
temperature $T=T_\xi $ at the crossover (see Eqs. \ref{ksi}, \ref{Lfai}). In
other words, the quasiparticle description holds over the {\it whole}
temperature range that corresponds to the WL regime\cite{aarev}. 

Since the
crossover temperature $T_\xi $ and the temperature $T_0$ pertinent to the
activation energy on the insulating side of the crossover are close to
one another, we do not distinguish between them\cite{power-law}. When
discussing the crossover temperature in this paper, we refer to $T_0$, which
can be accurately estimated on the insulating side of the crossover.

\begin{figure}[th]
\epsfig{file=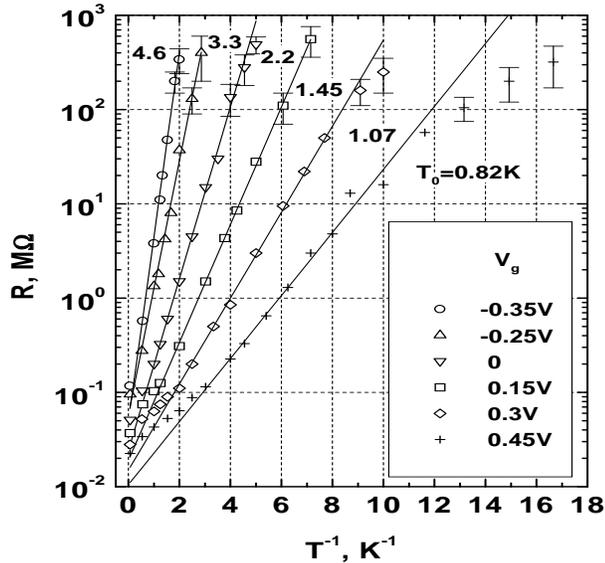, height=0.5\textwidth, width=0.5\textwidth}
\caption{The R(T) dependences at different values of the gate voltage for sample 1.
Straight lines are Arrhenius dependences (5) with the values of the activation energy $%
T_0$ shown next to the lines. }
\label{Fig.2}
\end{figure}

One can shift the crossover and vary the
activation energy over a wide range by applying the gate voltage 
(the temperature dependences of the
resistance for sample 1 at different $V_g$ are shown in Fig. 2). With
increasing the carrier concentration, the activation energy decreases (Fig. 3a)
and the localization length increases (the values of $\xi $
shown in Fig. 3b have been calculated from Eq.\ref{ksi}). However, the
product $T_0\xi $ remains independent of the gate voltage (Fig. 3c). Thus, the
activation energy is inversely proportional to $\xi $. Numerically, the
activation energy is very close to the mean energy spacing within the
localization domain; we have also verified this by studying samples of
different width (see Table 1).  

We observe correlation between the prefactor $R_0$ in the Arrhenius law (5)
and the sheet resistance $R_{\Box }$ on the metallic side of the crossover ($%
T>>T_0$). For example, the magnitude of $R_0$ calculated for a segment of
wire of the length $\xi $ ($R_{0\xi} $) is $1.7$ $k\Omega $ for sample 1 ($%
R_{\Box }(20K)=1.6k\Omega $) and $12-14$ $k\Omega $ for samples 3 and 5 ($%
R_{\Box }(20K)=3.5-4.2$ $k\Omega $). The presence of the gate electrode
affects $R_{0\xi} $: for sample 1, $R_{0\xi} $ has been increased up to $4$ $%
k\Omega $ after the gate deposition. With increasing the gate voltage, $R_0$
decreases (Fig. 3d), however, $R_{0\xi} $ is almost $V_g$-independent. This
observation can be also presented as the $V_g$-independent ratio 
$R_0/R_{\Box }(T>>T_0)$ for a given sample. As it will be shown below, the
prefactor $R_0$ is not affected by the magnetic field.

\begin{figure}[th]
\epsfig{file=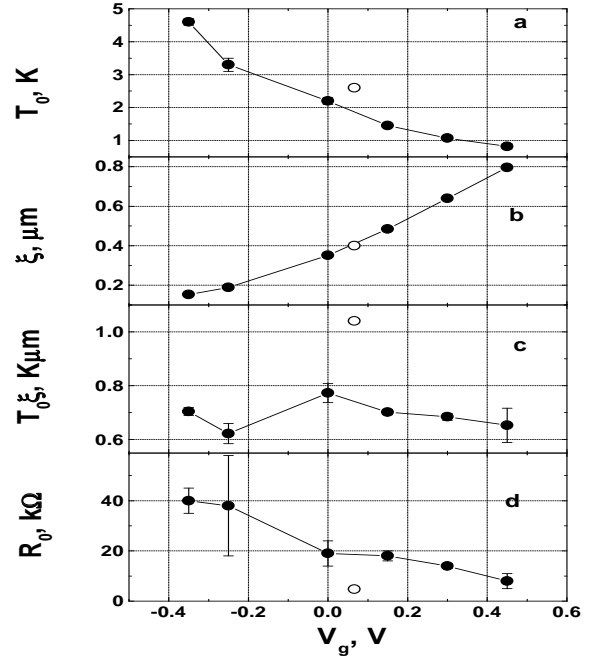, height=0.5\textwidth, width=0.5\textwidth}
\caption{The dependences of $T_0$, $\xi$, $T_0\xi$ and $R_0$
on the gate voltage for sample 1. Open dots are the corresponding 
values before the gate deposition. These values
have been plotted at non-zero $V_g$ to facilitate comparison of parameters
for the samples with/without a gate
with the {\it same} $\xi$. The values
of $\xi$ have been calculated from Eq. 1.}
\label{Fig.3}
\end{figure}

The observed Arrhenius-type temperature dependence of the resistance with $%
k_BT_0\approx \Delta _\xi $ could be accounted for by different models of
electron transport in the SL regime (see, f.i. \cite{kurk,lark-khmel,lee,raikh}). In
particular, the theory of the variable range hopping (VRH) in one dimension%
\cite{kurk,lee,raikh} predicts the activation behavior of the resistance (in
contrast to higher dimensions, where one can expect to observe either Mott's
or Efros-Shklovskii law\cite{es}). Similarity of the theoretical predictions stems
from the fact that the resistance of a Q1D conductor is governed by the
so-called {\it critical} hops, rare segments of a wire with no localized
states in the vicinity of the Fermi level\cite{lee,raikh}. Indeed, any
model of the SL transport that takes into account a realistic distribution
of parameters of the hops, brings to the highly-resistive hops separated by
a distance $L_c$ much larger than the hopping length. Thus, in order to test
the relevance of different theoretical models to our experimental situation,
we need to measure directly the two characteristic length scales: the
hopping distance  $r_h$ for the critical hops, and the distance  $L_c$
between such hops. The experimental data on the hopping distance is still unavailable;
without this information, it is difficult to distinguish between the alternatives:
nearest-neighbor hopping versus variable-range hopping.
We hope to address this problem in our future experiments with multiconnected
samples fabricated from Q1D wires. However, the second important length scale,
the distance between the critical hops, can be measured directly.

\subsection{Non-linearity of the I-V characteristics}

The study of non-linear
effects in the SL regime allows to measure $L_C$ and its dependence on $T$ and $H$
which is crucial for  understanding of electron transport in the SL
regime.
For all the samples in Table 1, we have measured the dependence of the
resistance $R\equiv V/I$ on the voltage $V$ across the sample at different
temperatures; we have also repeated these measurements for sample 1 after
the gate deposition for different $V_g$ (all the data discussed in Section
IIIA were obtained in the linear regime). The dependences $R(V)$ measured at
different $T$ and $H=0$ for sample 1 before the gate deposition are shown in
Fig. 4.

\begin{figure}[ht]
\epsfig{file=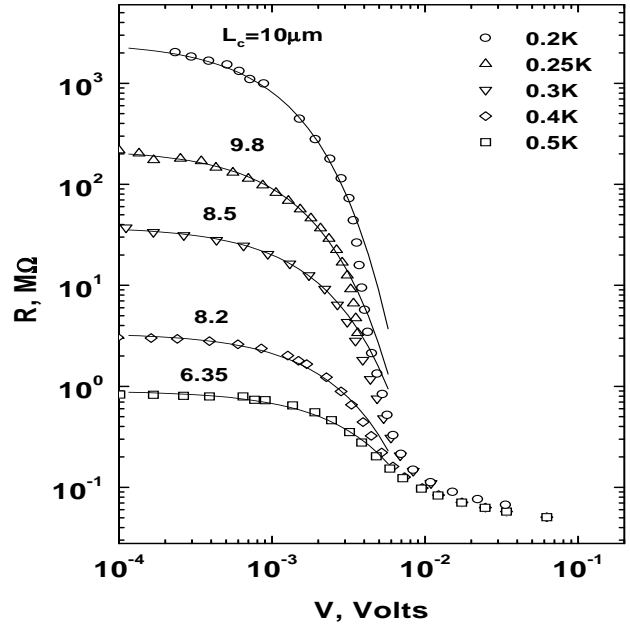, height=0.5\textwidth, width=0.5\textwidth}
\caption{The dependence of $R\equiv V/I$ on the voltage drop $V$ across 
sample 1 at different temperatures (before the gate deposition). Solid curves
are Eq. 7; the corresponding values of $L_C$ in microns are shown next to each curve.}
\label{Fig.4}
\end{figure}

Qualitatively, one can consider two different voltage regions for these
dependences, separated by the characteristic voltage $V^{*}$ ($\sim%
5\cdot 10^{-3}V$ for sample 1). At small $V<V^{*}$, the resistance is
strongly temperature-dependent; this voltage interval corresponds to the SL
regime. For large $V>V^{*}$, all dependences $R(V,T)$ collapse onto a single
curve regardless of the temperature; in the latter WL regime, the electron
transport is non-activated. Heating of sample 1 by measuring currents can
be neglected at $V<1\cdot 10^{-2}V$: independent measurements show that the
power $1nW$ dissipated in the sample does not overheat electrons down to $%
T\sim 0.1K$\cite{ref} . 

\begin{figure}[th]
\epsfig{file=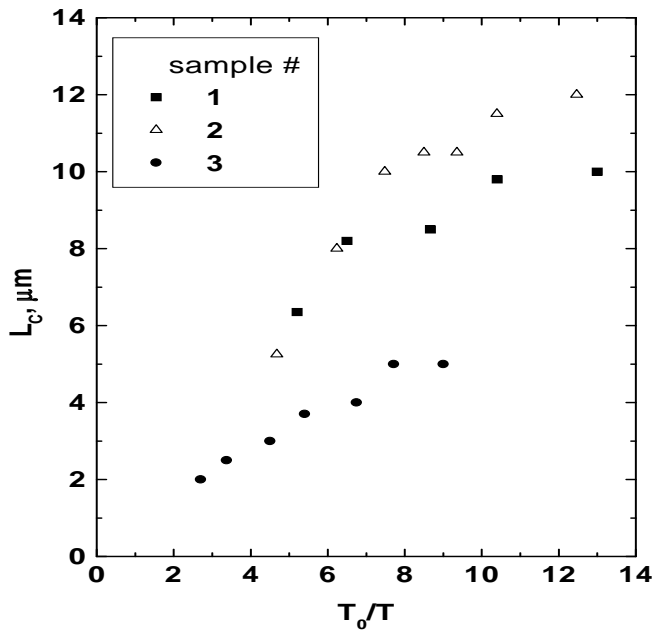, height=0.5\textwidth, width=0.5\textwidth}
\caption{The temperature dependence of $L_C$ for different samples; 
the sample parameters are given in Table 1.}
\label{Fig.5}
\end{figure}

In the SL regime, the non-linear resistance $R\equiv V/I$ for all samples
can be fitted with the dependence: 
\begin{equation}
R\sim \exp \left( \frac{T_0-\alpha V}T\right) \text{ ,}  \label{non-lin}
\end{equation}
where $V$ is the total voltage across the sample. In order to clarify the
physical meaning of $\alpha $, we assume the following simplified model: the
critical hops are identical, and they are separated by the average distance $%
L_C$ which is much greater than the hopping distance. In the electric field, 
the activation energy of each
critical hop decreases proportionally to the voltage drop across the hop.
In this model, $\alpha ^{-1}$ is proportional to the average number of critical hops
$L/L_C$ in a wire of the length $L$. 

Obviously, this model is very naive. However, more realistic model based on the
normal distribution of the activation energies and self-consistent calculation
of the voltage drop across each hop fits the experimental dependences $R(V)$ less accurately
than Eq. 7.

On the basis of this model, one can estimate the distance
between the critical hops $L_C=\alpha Lk_B/e.$ This distance increases with
decreasing the temperature (Fig. 5) and at $T_0/T\gg 1$ it exceeds $\xi $ by
more than an order of magnitude. However, even at the lowest temperatures,
this distance is still by a factor of $\sim 50$ smaller than the total
length of the wire. 

The theory \cite{raikh} predicts
that the wire-to-wire fluctuations can be neglicted if $\eta \equiv ln(L/\xi
)/ln(L_C/\xi )>1$; for our experimental values of $L_C$, $\eta =2\div 3$ for
all $T$ and $V_g$ for all samples. Since the width of the distribution function for the
wire's resistance depends exponentially on $\eta $, the wire-to-wire
fluctuations are averaged out in our ''long'' samples. Thus, the wire resistance 
is a {\it self-averaged} quantity in the studied temperature range (the resistance
fluctuations decrease with increasing the wire length). The opposite case of
large fluctuations in mesoscopic samples ($L\leq L_C$) has been studied by
Hughes {\it et al.}\cite{savch}.

\begin{figure}[ht]
\epsfig{file=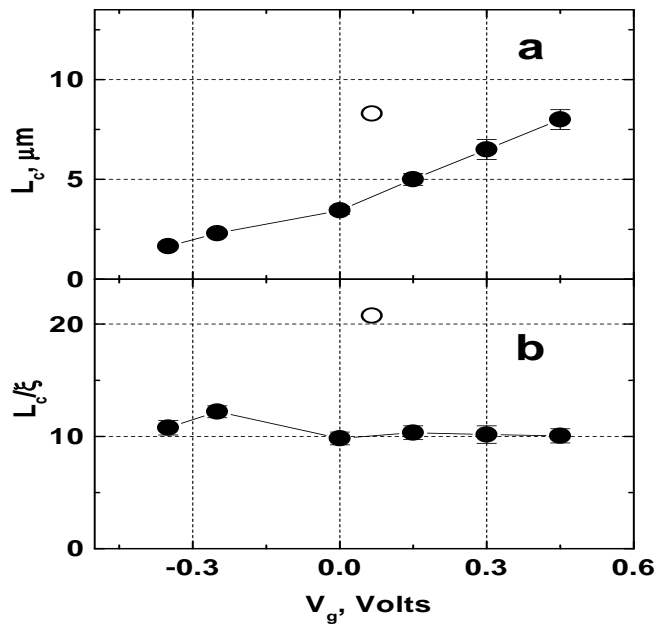, height=0.5\textwidth, width=0.5\textwidth}
\caption{a) The dependence of $L_C$ on $V_g$ at a fixed $T_0/T \simeq 8.25$.
b) The ratio $L_C/\xi$ versus $V_g$. The open circles
correspond to the measurements before the gate deposition.}
\label{Fig.6}
\end{figure}

Another experimental evidence for the self-averaged behavior of the wire resistance
stems from comparison of the samples comprising 5 wires with  $L=40\mu m$ and 
470 wires with $L=500\mu m$. For 40$\mu m$-long 
wires we observed resistance fluctuations in strong magnetic fields (Fig. 9; see also 
\cite{gersh1}); these fluctuations are completely washed out for longer wires (Fig. 8).
Relatively small
values of $L_C$ are also consistent with the fact that we have not observed
any rectifying effects even for the 40$\mu m$-long wires;
the rectifying effects are typical for mesoscopic samples \cite
{fow1}. 

The study of $L_C$ for the same sample at different $V_g$ 
shows that $L_C$ is {\it proportional} to $\xi $ for a fixed $T_0/T$. For sample 1 both $%
L_C$ and $\xi $ increase with increasing $V_g$ by a factor of $\sim 6$%
, however, the ratio $L_C/\xi $ remains the same for all $V_g$ (Fig. 6).

The temperature dependences of $L_C$ shown in Fig. 5 contradict 
the VRH theory in one dimension, which
predicts the exponential temperature dependence of $L_C$ \cite{lee,raikh}: 
\begin{equation}
L_C=\xi \sqrt{\frac{T_0}T}\exp \left( \frac{T_0}{2T}\right) \text{ .}
\end{equation}
Instead, $L_C$ grows approximately as $T_0/T$ at high temperatures and saturates at
lower $T$ (Fig. 5). This discrepency remains the challenge for the theory.

\subsection{The magnetoresistance}

An important advantage of our experiment is that we can measure directly
the localization length by studying the magnetoresistance in the SL regime.
The magnetoresistance for sample 1 below the crossover temperature is shown
in Fig. 7. The magnetoresistance in the
WL and SL regimes shares several common features: it is negative and strongly
anisotropic (this pure orbital
magnetoresistance vanishes for the parallel orientation of the field with
respect to the plane of the $\delta $-layer). However, the magnitude of the magnetoresistance
increases dramatically on the insulating side of the crossover. The insert
in Fig. 7 demonstrates that the crossover shifts down to lower 
temperatures, and the magnetoresistance  
becomes exponentially strong in classically weak magnetic fields. 

\begin{figure}[th]
\epsfig{file=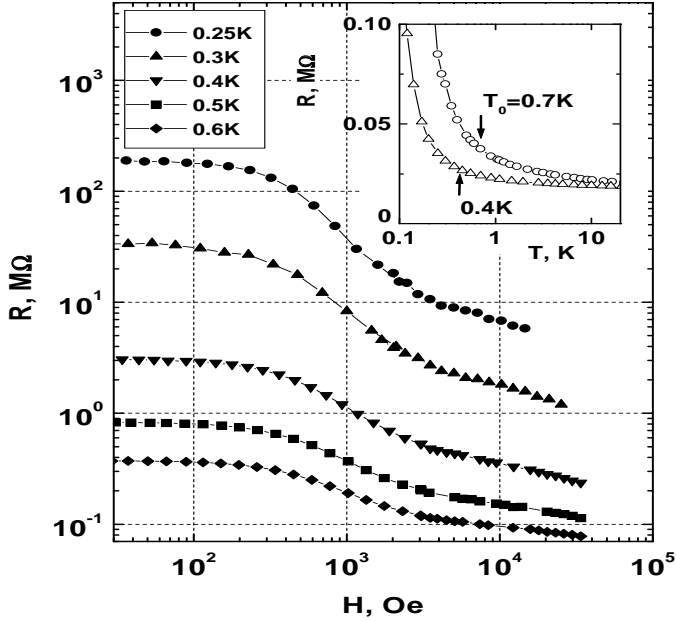, height=0.5\textwidth, width=0.5\textwidth}
\caption{The magnetoresistance of sample 1 without the gate electrode
 at different temperatures ($T\ll T_0=2.6K$). The solid lines are
guides to the eye. The insert: the shift of the crossover in the
magnetic field for the same sample with the gate electrode at $V_g=+0.7V$
($\bigcirc - H=0, \bigtriangleup - H=17kOe $). }
\label{Fig.7}
\end{figure}

For different fixed values of the magnetic field, we observe the exponential
temperature dependence of the resistance with the same prefactor $R_0$.
The {\it only} parameter which varies with the magnetic field is the
activation energy:

\begin{equation}
R(T,H)=R_0exp[T_0(H)/T] \text{ .} \label{R(H)}
\end{equation}
Thus, the magnetoresistance is due to the magnetic field dependence of the
activation energy\cite{gersh1,gersh2}. Taking this into account, it is
convenient to convert the magnetoresistance into the $H$-dependence of the
activation energy:

\begin{equation}
%TCIMACRO{\QOVERD. . {T_0(H)}{T_0(H=0)}}
%BeginExpansion
{T_0(H) \overwithdelims.. T_0(H=0)}
%EndExpansion
=%
%TCIMACRO{\QOVERD. . {T}{T_0(H=0)}}
%BeginExpansion
{T \overwithdelims.. T_0(H=0)}
%EndExpansion
\ln 
%TCIMACRO{\QOVERD. . {R(H)}{R_0}}
%BeginExpansion
{R(H) \overwithdelims.. R_0}
%EndExpansion
.  \label{T0H}
\end{equation}
The dependences (\ref{T0H}) measured for samples 1 and 5 at different temperatures $%
T\ll T_0$ are shown in Figs. 8,9. For all the samples, these dependences
collapse onto the{\it \ same universal} curve, which reflects the transition
from weak to strong fields; the normalized activation energy varies between $%
1$ ($H=0$) and $\sim0.5$ (strong fields). (For samples with $L=40\mu
m$, the deviations from this universal curve due to insufficient averaging
of mesoscopic fluctuations have been observed in strong fields\cite{gersh1},
see Fig.9). 

It was shown in Sec.IIIA that the activation energy in our samples practically coincides
with the mean energy spacing within the localization domain and is inversely
proportional to the localization length. Thus, the observed 
magnetoresistance reflects the universal magnetic-field dependence of the
localization length.

This observation is in agreement with the theoretical prediction that the
localization length in a 1D conductor with a large number of channels $N_{1D}$
should be a universal function of the symmetry class\cite{larkin,pichard,fyodorov,been}:

\begin{equation}
\xi =\gamma N_{1D}l\text{ ,} \label{mirlinksi}
\end{equation}
where $\gamma =2\beta /s$, $\beta $ is the Dyson parameter, which
characterizes the symmetry properties of the system, and $s$ is the Kramers
degeneracy of the channels. The coefficient $\gamma $ equals $1(2,4,4),$
respectively, for potential scattering ($\beta =1$, $s=2)$, potential
scattering in strong magnetic field ($\beta =2$, $s=2$), spin-flip
scattering by magnetic impurities {\it or} the strong spin-orbit 
scattering in strong magnetic field ($\beta =2$, $s=1$, note broken Kramers
degeneracy), and the strong SO scattering at $H=0$ ($\beta =4$, $s=2$). In
our case, the magnetic field induces a transition from the orthogonal to
unitary case without breaking the spin degeneracy of the scattering channels
($\beta =1$, $s=2\rightarrow \beta =2$, $s=2$) and, hence, {\it doubling} of $\xi $%
. The theory is well adapted to the conductors with a large localization
length, where electrons move diffusively within the localization domain.
Since for our samples $T_0\simeq \Delta _\xi \sim 1/\xi ,$ {\it doubling} of
the localization length should result in {\it halving} the activation
energy in agreement with our experimental data. 

\begin{figure}[ht]
\epsfig{file=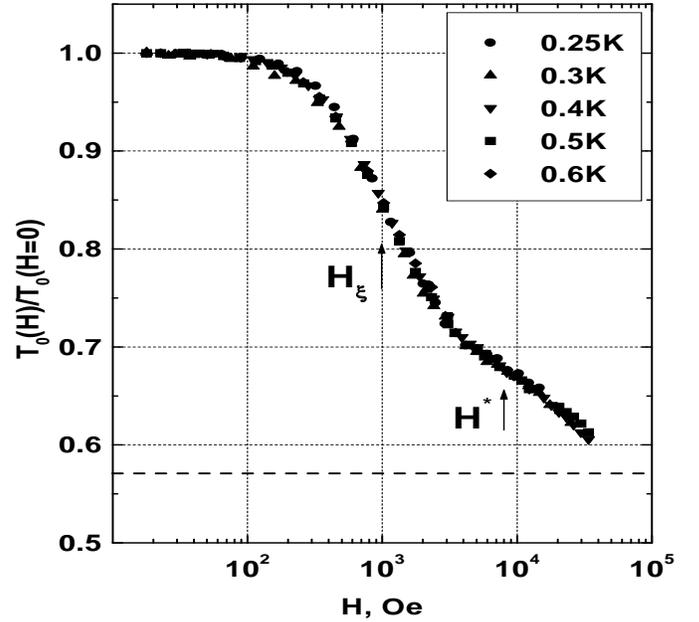, height=0.5\textwidth, width=0.5\textwidth}
\caption{The normalized magnetic field dependences of the activation
energy for sample 1. 
Characteristic fields $H_\xi$ and $H^{*}$ are shown with arrows. The dashed line
is the theoretical strong-field limit for $N_{1D}=7$ (see the text).}
\label{Fig.8}
\end{figure}

Figure 8 shows that the activation energy for sample 1 diminishes in strong
magnetic fields less than by a factor of 2. We believe that two reasons
preclude observation of the {\it exact} halving of $T_0$ in this sample.
First, the number of channels is not very large for sample 1 ($N_{1D}\simeq 7$); in this case,
the exact expression for $\xi $ should be used $(s=2)$\cite{been}: 
\begin{equation}
\xi =\left( \beta N_{1D}+2-\beta \right) l\text{ ,}  \label{ksiH}
\end{equation}
According to Eq.\ref{ksiH}, for a conductor with $N_{1D} =7$, $T_0$ in
strong fields should be smaller than $T_0(H=0)$ by a factor of $1.75$. The
corresponding high-field limit of $T_0$ is shown as the dashed line in Fig.
8. Note that for sample 5 with a large number of transverse channels ($N_{1D}\simeq 30$), the
normalized $T_0$ approachs the level 0.5 in strong fields (Fig. 9). 
Secondly, in stronger fields $H>H^{*}=\Phi _0/W^2$($\Phi _0$ is the magnetic
flux quantum), two essential requirements of the theory applicability are
violated: sample 1 becomes two-dimensional with respect to the localization
effects, and, at the same time, transport is already affected by the
magnetic field at scales smaller than the mean free path (since $l$ is close
to $W$ for this sample). In this case, the dependence $\xi \left( \beta \right) $ is
more complicated and not universal\cite{bouchaud,lerner}. 

\begin{figure}[ht]
\epsfig{file=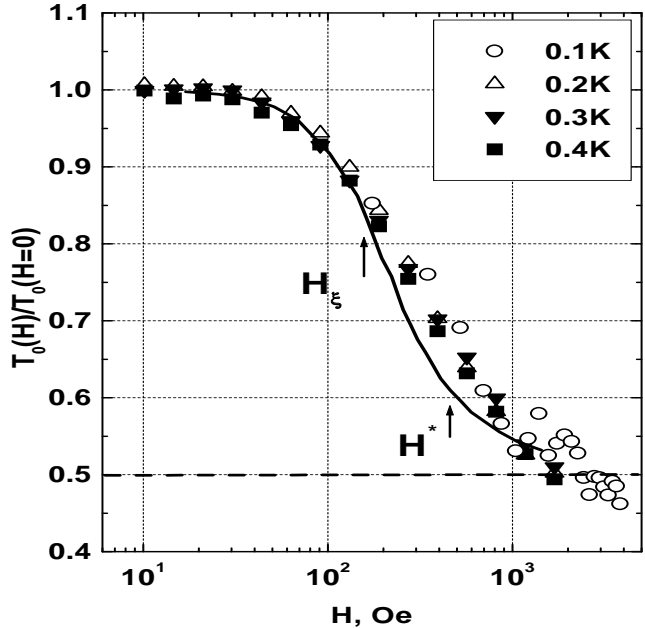, height=0.5\textwidth, width=0.5\textwidth}
\caption{The normalized magnetic field dependences of the activation
energy for sample 5. The solid line is the theoretical dependence
$\xi(0)/\xi(H)$ ($=T_0(H)/T_0(0)$) for $N_{1D}\gg 1$ (see the text). The dashed line
is the theoretical strong-field limit. The amplitude of reproducible
fluctuations of magnetoresistance, observed at $H\geq 1kOe$, increases
with the decrease of temperature; the fluctuations clearly manifest
themself at $T=0.1K$.}
\label{Fig.9}
\end{figure}

Not only the limits of variation of $\xi (H)$ are in agreement with the
theory, but also the shape of the transition curve in the field range $H$ $%
<H^{*}$ is consistent with the analytical expression for $\xi (H)$ obtained
for all magnetic fields by Bouchaud\cite{bouchaud}. The dependence $\xi (H)$
calculated for $N_{1D}\gg 1$ is shown in Fig. 9 by the solid curve. In
particular, our data are consistent with the prediction that the limit $\xi
(H)/\xi (0)$ is reached slowly with increasing $H$\cite{bouchaud}.

Our experiments \cite{gersh1,gersh2,gersh3} provide the first evidence of
the doubling of $\xi $ due to breaking of the time-reversal symmetry.
Previously, the idea of the universal change of $\xi $ in magnetic fields
has been used for interpretation of the magnetoresistance of several 2D and
3D systems with variable-range hopping\cite{pichard,hernandes,ladieu}.
Although the effects in higher dimensions could be qualitatively similar,
the doubling of $\xi $ is expected only in the Q1D geometry\cite
{bouchaud,lerner}. It is unclear at present how to reconcile the observed
positive magnetoresistance in insulating 1D samples with strong SO scattering\cite
{hernandes}, with the theory that properly accounts
for Kramers degeneracy\cite{meir,mirlin}. In this case, the magnetic field
should not affect the localization length, since the time-reversal symmetry
and Kramers degeneracy are broken simultaneously\cite{larkin,meir,mirlin}.

Observation of the magnetic-field-induced doubling of $\xi $ provide us with
a {\it direct} method of measurement of $\xi $ in Q1D conductors. Indeed,
the localization length is the only unknown parameter in fitting the
experimental dependences $T_0(H)/T_0$ with the theory\cite{bouchaud}. According to the
theory\cite{bouchaud}, $T_0(H)/T_0(H=0)\approx 0.83$ in the characteristic
field 
\begin{equation}
H_\xi =\frac{\Phi _0}{\xi W}  \label{Hksi}
\end{equation}
which corresponds to breaking of the time-reversal symmetry within the
localization domain\cite{Hksi}. The value of $H_\xi $ for samples 1 and 5 are
shown by arrows in Figs. 8,9. For all the samples studied, the experimental
values of $\xi $ are in an excellent agreement with the estimate of the
localization length from the resistance in the ''metallic'' regime (\ref{ksi})
(see Table 1).

\begin{figure}[ht]
\epsfig{file=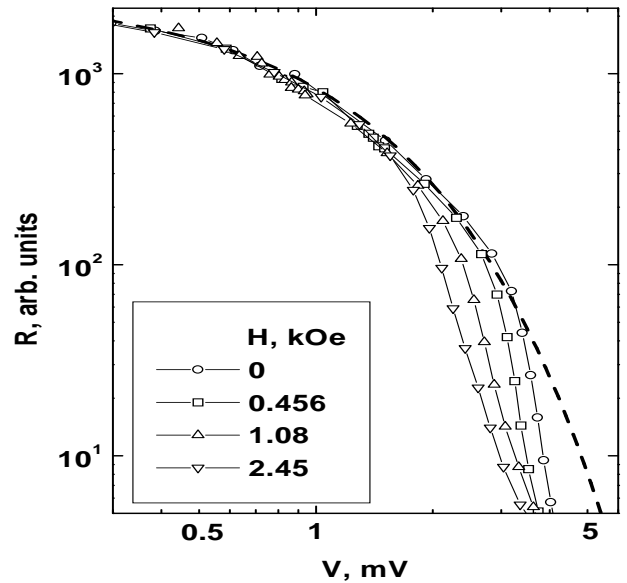, height=0.5\textwidth, width=0.5\textwidth}
\caption{The normalized $R(V)$ dependences at different $H$, 
 $T=0.2K$. Solid lines are guides to the eye, the dashed line -
Eq. 7. }
\label{Fig.10}
\end{figure}

The evolution of the $R(V)$ dependences with magnetic field for sample 1 is
shown in Fig. 10. In the SL regime ($V<<V^{*}$), these normalized dependences
are not affected by the magnetic field in accord with our experimental observation
that $L_C\sim \xi
(T_0/T)=1/\nu _{1D}T$. However, the values of $V$, where deviation from the fitting
curve (7) is observed, are diminishing with the increase of magnetic field.
We cannot suggest a plausible explanation of this experimental fact.

\subsection{The density of states}

Comparison of expressions for the activation energy $k_BT_0$ $\approx \Delta _\xi $ and the
characteristic field $H_\xi $ shows that the ratio of these quantities
 depends only on the single-particle density of states: 
\begin{equation}
\frac{H_\xi }{\Delta _\xi }=%
%TCIMACRO{\QOVERD. . {\Phi _0\nu _{1D}}{W}}
%BeginExpansion
{\Phi _0\nu _{1D} \overwithdelims.. W}
%EndExpansion
=\Phi _0\nu _{2D} \text{ .} \label{DOS}
\end{equation}
Thus, by measuring the experimental counterpart of this ratio, $H_\xi
^{exp}/T_0$, one can probe the DoS 
at the energy scale $\sim k_BT_0$ near the Fermi level.
The values of $H_\xi ^{\exp }/T_0$ for samples without a gate are listed in
Table 1. Despite of an order-of-magnitude variation of $H_\xi $ and $T_0$
for different samples, their ratio remains close to the estimate 0.5 kOe/K,
obtained for the non-interacting electrons in the parabolic conduction
 band ($\nu _{2D}=m^{*}/\pi \hbar ^2$ , $m^{*}$=0.067$m_e$). 

\begin{figure}[ht]
\epsfig{file=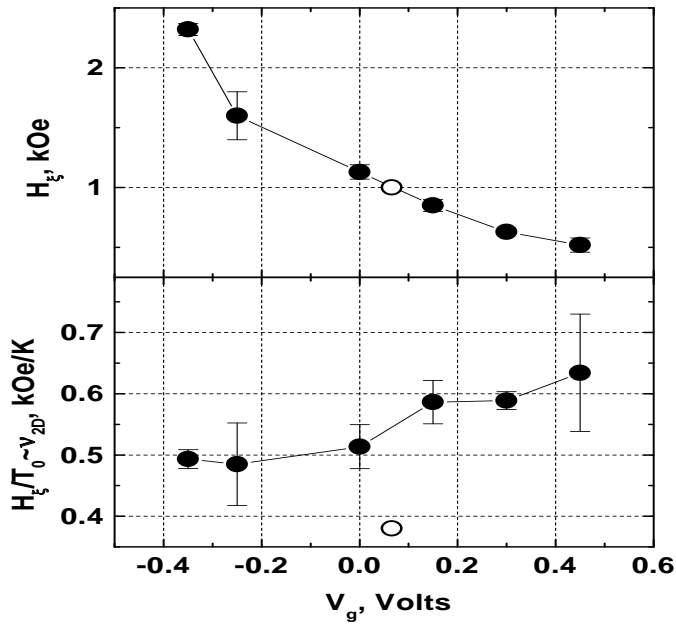, height=0.5\textwidth, width=0.5\textwidth}
\caption{The dependence of $H_\xi$ and $\nu_{2D}$ on $V_g$. Solid lines are
guides to the eye. Open circles correspond to the values before the gate
deposition. }
\label{Fig.11}
\end{figure}

Interestingly, however, we observe $\sim$ 40\% increase of $\nu_{2D}$ after 
{\it deposition} of the gate electrode (Fig. 11). The increase of $\nu _{2D}$
also manifests itself in the decrease of $T_0$ (Fig. 3) and increase of 
$L_C$ (Fig. 6) if one compares the samples with the same localization length.
This behavior of $\nu _{2D}$ is difficult to explain in the model of
non-interacting electrons. We believe that this is manifestation of the
effect of the {\it long-range} Coulomb interaction on the DoS in the SL
regime. Indeed, the Coulomb interaction, being poorly screened in Q1D
conductors, can affect both thermodynamic and transport properties. In
particular, Raikh and Efros predicted a logarithmic singularity of the
single-particle DoS at the Fermi energy in Q1D conductors in the SL regime\cite{raikefr}:

\begin{equation}
\nu _{1D}(\varepsilon )=%
%TCIMACRO{
%\QOVERD. . {\nu _{1D}^0}{1+\QOVERD. . {e^2\nu _{1D}^0}{K}\ln \left( \varepsilon _\xi /\varepsilon \right) }}
%BeginExpansion
{\nu _{1D}^0 \overwithdelims.. 1+{e^2\nu _{1D}^0 \overwithdelims.. K}\ln \left( \varepsilon _\xi /\varepsilon \right) }
%EndExpansion
\text{ ,}  \label{raikhDoS}
\end{equation}
where $\nu _{1D}^0$ is the ''nucleating'' DoS, $K$ is the relative
permittivity of the medium around the conductor, and $\epsilon_\xi$
gives a measure of the strength of nearest-neighbor interaction.
Suppression of the long-range Coulomb interaction can result in the
measurable increase of $\nu _{2D}$. Indeed, deposition of the gate electrode
''screens'' the long-range part of the Coulomb interaction: it becomes of a
dipole-dipole type at distances greater than the distance $t$ between the
electron gas and the metallic gate electrode (for our samples, $%
t=0.1\mu m$ is much smaller than $\xi $ and the hopping length). To the best
of our knowledge, this is the first experimental evidence of the minimum of
the DoS at the Fermi level in Q1D conductors. This could also be an indirect
evidence that the hopping distance at $T\ll T_0$ exceeds $\xi$: only in this case 
one can expect to observe the effect of the Coulomb interaction on the DoS.

\section{CONCLUSION}

In conclusion, we have studied the resistance of quasi-one-dimensional wires fabricated
from Si $\delta $-doped GaAs as a function of the temperature, magnetic
field and applied voltage. The crossover from weak to strong localization
has been observed in these conductors with decreasing the temperature.
The main features of the observed crossover, driven by both localization and
interaction effects, are in agreement with the Thouless scenario: the
crossover occurs when the phase-breaking length becomes comparable with the
localization length, and the resistance of the segment of wire of the length 
$\xi $ is $\sim h/e^2.$ 

On the insulating side of the crossover, we observe the activation
temperature dependence of the resistance with the activation energy very
close to the mean energy spacing within the localization domain. Both the
crossover temperature and the activation energy can be varied over a wide
range by the gate voltage. 

The study of non-linearity of the current-voltage characteristics in the SL
regime provides the direct measurement of the distance between the critical
electron hops, which govern the resistance of a Q1D conductor. This distance 
$L_c$ is proportional to the localization length; it increases
with decreasing the temperature, and at low temperatures ($T/T_0=0.1)$
exceedes $\xi $ by a factor of $\sim 30$. However, $L_c$ is insufficiently
large to be consistent with prediction of the theory of variable range hopping
in Q1D conductors\cite{lee,raikh}. Since $L_c$ is much smaller than the length of wires,
there is no rectifying effects in the resistance of our samples, and the
wire-to-wire fluctuations of the resistance are negligible.

The exponentially strong magnetoresistance in the SL regime is due to the
magnetic field dependence of the localization length. Observation of the
universal magnetic-field dependence of the activation energy, which is
caused by breaking of the time-reversal symmetry in strong fields, has been used
for the direct measurement of $\xi $ in Q1D conductors. There is a good
agreement between the values of $\xi $ estimated from the SL
magnetoresistance and calculated from the resistance in the WL regime. 

Simultaneous measurement of the activation energy and the characteristic
field of doubling of the localization length allows to probe the single-particle 
density of states at the Fermi level in Q1D conductors. Our data indicate, that
deposition of the gate electrode decreases the amplitude of the minimum. We
believe that this is due to screening of the long-range interaction by the
metal film separated from the Q1D conductor by a distance much smaller than
the hopping length.

In the situation when direct measurement of the hopping distance is still 
unavailable, it is difficult to give preference to one of the models of 
electron transport in the insulating regime (nearest-neighbor hopping
versus variable-range hopping). However, observation of the minimun of
the density of states at the Fermi level can serve as an indirect evidence 
that the hopping distance exceeds the localization length.
More theoretical efforts are
needed to take into account such essential features of quasi-one-dimensional
conductors as strong
overlapping between the localized electron states and long-range Coulomb
interaction. 

\section{Acknowledgements}

We are grateful to our collaborators, A. G. Mikhalchuk and H. M. Bozler;
their help was essential at the initial stage of this work. It is our
pleasure to thank I. L. Aleiner, B. L. Altshuler, A. I. Larkin, A. D.
Mirlin, M. E. Raikh, and B. I. Shklovskii for helpful discussions. We are
grateful to B. K. Medvedev for fabrication of the $\delta $-doped GaAs
layers.


\begin{references}
\bibitem{power-law}  Because all the relevant energies are of the same
magnitude at the crossover, the {\it power-law} temperature dependence of
the conductivity, $\sigma \sim \tau _\varphi ^{-1}\sim T^p$, that has been
suggested for the temperature range $T_0<T<T_\xi $ \cite{imry,thorn}
cannot be observed in 1D conductors with the dominant Nyquist phase-breaking.
\bibitem{ref}  The resistance of the GaAs layer between the gate electrode
and the electron gas in our samples is strongly non-linear: for sample 1, this
resistance, $R_g$, is very large at $V_g<0.5V$, however, starting with $%
V_g\sim 0.5V$, it decreases with increasing $V_g$ down to 0.5 G$\Omega $ at $%
V_g=2V$. For this reason, all the measurements have been performed at $%
V_g<0.5V$. The power $P=V_g^2/R_g$, which is dissipated in the sample, is 1
nW at $V_g=1V$; this power does not affect the electron temperature down to
~0.1 K.
\bibitem{Hksi}  In our previous papers \cite{gersh2,gersh3}, we extracted
the values of $H_\xi $ by fixing the level $T_0(H_\xi )=0.85T_0(0)$ (at that
time we were unaware of the calculations \cite{bouchaud}). This choice
was based on comparison of the experimental data for many samples with the
theoretical estimate for $H_\xi $. The calculated values of $H_\xi $ were in
an excellent agreement with the experimental ones; this indicates that the
shape of the transition curve is the same for all the samples.
\end{references}
\end{document}